\documentclass[journal]{IEEEtran}

\usepackage{cite}
\usepackage{amsmath,amssymb,amsfonts}
\usepackage{algorithmic}
\usepackage{graphicx}
\usepackage{textcomp}
\def\BibTeX{{\rm B\kern-.05em{\sc i\kern-.025em b}\kern-.08em
    T\kern-.1667em\lower.7ex\hbox{E}\kern-.125emX}}

\usepackage[utf8]{inputenc}
\usepackage[english]{babel}
\usepackage{booktabs}
\usepackage{multirow}

\usepackage{balance}
\usepackage{algorithm}

\newcommand{\lora}{\textsc{LoRa}}
\newcommand{\lorawan}{\textsc{LoRaWAN}}
\newcommand{\sigfox}{\textsc{SigFox}}

\newcommand{\zbn}{\textsc{IEEE~802.15.4g}}
\newcommand{\wsun}{\textsc{Wi-SUN}} 

\begin{document}

\title{Optimum \lorawan{} Configuration\\Under \wsun{} Interference}

\author{Arliones~Hoeller, Richard~Demo~Souza, Hirley~Alves, Onel L. Alcaraz López,\\Samuel Montejo-Sánchez, Marcelo Eduardo Pellenz%
\thanks{A. Hoeller and R.D. Souza are with the Department of Electrical and Electronics Engineering, Federal University of Santa Catarina, Florianópolis, 88040-900 Brazil.}
\thanks{A. Hoeller is also with the Department of Telecommunications Engineering, Federal Institute for Education, Science, and Technology of Santa Catarina, São José, 88103-310 Brazil.}
\thanks{A. Hoeller, H. Alves, and O.L.A. López are with Centre for Wireless Communications, University of Oulu, Oulu, 90014 Finland.}
\thanks{S. Montejo-Sánchez is with Programa Institucional de Fomento a la I+D+i, Universidad Tecnológica Metropolitana, Santiago, 8940577 Chile.}
\thanks{M.E. Pellenz is with the Informatics Graduate Program, Pontifical Catholic University of Paraná, Curitiba, 80215-901 Brazil.}
\thanks{Correspondence author: Arliones.Hoeller@ifsc.edu.br.}
\thanks{This work has been partially supported in Brazil by the National Council for Scientific and Technological Development (CNPq), Print CAPES-UFSC ``Automation 4.0'', and INESC P\&D Brasil (Project F-LOCO,  Energisa, ANEEL PD-00405-1804/2018); in Finland by Academy of Finland (Aka) 6Genesis Flagship (Grant 318927), EE-IoT (Grant 319008) and Aka Prof (Grant 307492); and in Chile by FONDECYT Postdoctoral (Grant 3170021).}
\thanks{\textcopyright 2019 IEEE. Personal use of this material is permitted. Permission from IEEE must be obtained for all other uses, in any current or future media, including reprinting/republishing this material for advertising or promotional purposes, creating new collective works, for resale or redistribution to servers or lists, or reuse of any copyrighted component of this work in other works.}}

\maketitle

\begin{abstract}
Smart Utility Networks (SUN) rely on the Wireless-SUN (\wsun{}) specification for years.
Recently practitioners and researchers have considered Low-Power Wide-Area Networks (LPWAN) like \lorawan{} for SUN applications.
With distinct technologies deployed in the same area and sharing unlicensed bands, one can expect these networks to interfere with one another.
This paper builds over a \lorawan{} model to optimize network parameters while accounting for inter-technology interference.
Our analytic model accounts for the interference \lorawan{} receives from \zbn{} networks, which forms the bottom layers of \wsun{} systems.
We derive closed-form equations for the expected reliability of \lorawan{} in such scenarios.
We set the model parameters with data from real measurements of the interplay among the technologies.
Finally, we propose two optimization algorithms to determine the best LoRaWAN configurations, given a targeted minimum reliability level.
The algorithms maximize either communication range or the number of users given constraints on the minimum number of users, minimum communication range, and minimum reliability.
We validate the models and algorithms through numerical analysis and simulations.
The proposed methods are useful tools for planning interference-limited networks with requirements of minimum reliability.

\end{abstract}

\section{Introduction}
\label{sec:intro}

Smart Utility Networks (SUN) are key enablers of Smart Cities~\cite{Centenaro:IEEEWC:2016}.
In such environments, the Internet-of-Things (IoT) plays a paramount role in connecting massive numbers of devices like smart meters, smart light bulbs, and smart appliances.
Besides the existence of several potential network technologies, \textit{e.g.}, \lorawan{}, \sigfox{}, and \wsun{}, the reliable and efficient connection of massive numbers of devices is still a challenge~\cite{Bockelmann:IEEECM:2016}.

Industry backs two initiatives: \textsc{LoRa Alliance} and \textsc{Wi-SUN Alliance}.
\textsc{LoRa Alliance} -- supported by Semtech, IBM, Cisco, Orange, among others -- maintains the \lorawan{} specification~\cite{LoRaAlliance:Site}.
\lorawan{} is a Low-Power Wide-Area Network (LPWAN) technology operating in the sub-GHz ISM band, using chirp spread-spectrum modulation, allowing increased signal robustness and range at low power consumption and low data rates~\cite{Centenaro:IEEEWC:2016}.
\lorawan{} uses the \lora{} physical layer (PHY) designed by Semtech and specifies the upper layer protocols to enable IoT deployments.
\textsc{Wi-SUN Alliance} -- supported by Cisco, Analog Devices, Toshiba, and others -- maintains the \wsun{} specification~\cite{WiSUN:site,Chang:2012}.
\wsun{} is a Field Area Network  (FAN) technology built upon the physical and link layers defined by the \zbn{} standard.
\zbn{} operates in different ISM bands, including the same sub-GHz bands used by \lorawan{}, where it employs a Gaussian Frequency Shift Keying (GFSK) modulation over narrow-band channels.
\wsun{} also defines network- and application-level services and profiles for different utility applications (e.g., energy, gas, water).

While utility service providers modernize their systems to use smart meters, deployments can use different communication technologies in the same geographical region, thus raising the question of how inter-technology interference affects network scalability.
Coordination among transmissions in different technologies is unfeasible at the network or lower layers, so it is essential to understand the impact of external interference through use cases, theoretical analysis, and design strategies~\cite{Zhang:WC:2019}.
To achieve a realistic model, one should take into account that LPWAN devices in the ISM radio band are subject to interference generated by other networks sharing the same part of the spectrum.
For instance, different authors report the analysis of the interaction of \lora{} with other technologies considering \zbn{}~\cite{Orfanidis:WiMob:2017}, \sigfox{}~\cite{Krupka:ME:2016,Poorter:WPC:2017}, and \textsc{IEEE 802.11ah}~\cite{Poorter:WPC:2017}.
The results in those papers suggest that \lora{} susceptibility to interference arriving from other technologies depends not only on the activity on those interfering networks but also on the configuration of the \lora{} signal, mainly the spreading factor (SF).
In this paper, we consider \lorawan{} as our target technology and model its performance in the presence of \zbn{} interference sources in the sub-GHz ISM band (e.g., around $868$~MHz in Europe and $915$~MHz in USA/Brazil).
Please note that the restriction of the model to \zbn{} interference comes without loss of generality since one can extend it to other network technologies provided that appropriate isolation thresholds between the technologies are available.

In our work, we evolve from previous developments in~\cite{Hoeller:Access:2018} and~\cite{Mahmood:2019} to approach the problem from an analytic perspective.
We derive two optimization algorithms that explore the configuration space of \lorawan{} in the presence of internal and external interference.
The algorithms are network planning tools for massive IoT applications, guiding the trading-off between reliability, the number of users, and coverage area/range.
We validate our analytic findings with simulations configured according to experimental results on the interplay of these networks published in~\cite{Orfanidis:WiMob:2017}.
We do not consider latency in this paper because our methods do not impact latency.
A good third-party work that analyzes the latency of Class A \lorawan{} is~\cite{Sorensen:WCL:2017}.

The contributions of this work include a closed-form expression for the inter-SF \lorawan{} interference model of~\cite{Mahmood:2019}; the extension of the analytic models of~\cite{Hoeller:Access:2018} and~\cite{Mahmood:2019} to consider external interference; the performance analysis of \lorawan{} considering the experimental results on inter-technology interference from~\cite{Orfanidis:WiMob:2017}; and two algorithms to optimize  \lorawan{} configuration, either in terms of network load or communication range, under reliability constraints.

The remaining of this paper is organized as follows.
Section~\ref{sec:related} summarizes related work, and Section~\ref{sec:lpwan} briefly introduces the characteristics of \lorawan{} and \zbn{}.
Section~\ref{sec:model} introduces the proposed models.
Section~\ref{sec:reliability} presents the proposed algorithms.
Section~\ref{sec:results} evaluates the models and algorithms.
Section~\ref{sec:conclusion} concludes the paper.

\section{Related Work}
\label{sec:related}

Georgiou and Raza~\cite{Georgiou:WCL:2017} propose an analytic model of \lorawan{}, which considers both disconnection and collision probabilities in Rayleigh fading channels. They show that \lorawan{} is sensitive to node density because it affects collision probability.
In~\cite{Hoeller:Access:2018}, we extend the work of~\cite{Georgiou:WCL:2017} to exploit message replications and multiple receive antennas at the gateway. We show that message replication is an interesting option for low-density networks, while the performance gains from spatial diversity are significant in all cases.
Mahmood \textit{et al.}~\cite{Mahmood:2019}, as well as we~\cite{Hoeller:Access:2018} and Georgiou and Raza~\cite{Georgiou:WCL:2017}, use stochastic geometry and Poisson Point Processes (PPP) to derive analytic models of the \lorawan{} coverage probability.
Contrasting with~\cite{Georgiou:WCL:2017} and~\cite{Hoeller:Access:2018}, the work in \cite{Mahmood:2019} considers the effect of interference from the imperfect orthogonality of \lora{} signals with different SF.

Orfanidis \textit{et al.}~\cite{Orfanidis:WiMob:2017} report an experimental evaluation of the interference between \zbn{} and \lora{} by measuring the packet error ratio in different SINR scenarios inside an anechoic chamber. The measurements consider a single \zbn{} interferer over one \lora{} link. Their results show that lower SFs are more susceptible to interference.
Although these measurements show interesting results, it is important to note that they consider a limited number of nodes, thus making it hard to extrapolate the conclusions.
To the best of our knowledge, no other work has studied the susceptibility of \lora{} to external \zbn{} interference, and there is none published work that investigates this relationship in a network-scale scenario with several active links in both \zbn{} and \lorawan{}.

\section{LPWAN Networks}
\label{sec:lpwan}

LPWAN technologies employ low-power communication mechanisms to enable the connection of thousands of IoT devices.
Most technologies work in sub-GHz frequencies and feature link budgets of $150\pm10$~dB, implementing robust communication channels with low energy consumption reaching distances in the order of kilometers~\cite{Centenaro:IEEEWC:2016}.
For reducing complexity and energy consumption, LPWANs use MAC protocols that may decrease channel usage efficiency. 
For instance, the unslotted ALOHA MAC in \lorawan{} presents high collision probability with large numbers of users~\cite{Abramson:FJCC:1970}.

\subsection{\lorawan{}} 

\lora{} is a proprietary sub-GHz chirp spread spectrum modulation technique optimized for long-range low-power applications at low data rates~\cite{LoRaAlliance:Site}.
\lora{} modulation depends, basically, on three parameters~\cite{Semtech:2015}: bandwidth ($B$), usually set to $125~$kHz or $250~$kHz for uplink and $500~$kHz for downlink; SF, which assumes values from 7 up to 12; and the forward error correction  (FEC) rate, varying from $\frac{4}{8}$ to $\frac{4}{5}$.
The parameters allow computing packet Time-on-Air (ToA), receiver sensitivity, and required signal-to-noise ratio (SNR) for successful detection in the absence of interference, which Table~\ref{tab:lora_specs} presents for a given packet configuration.
Note that ToA grows exponentially with SF, reducing the data rate and decreasing receiver sensitivity, improving coverage.

\begin{table}[ht]
\centering
\caption{\lora{} Uplink characteristics for 9-byte packets, $B = 125$~kHz, CRC and Header Mode enabled, and FEC rate $\frac{4}{5}$ for the SX1272 transceiver~\cite{Semtech:SX1272:2015}.}
\label{tab:lora_specs}
\begin{tabular}{@{}ccccc@{}}
\toprule
\textbf{\begin{tabular}[c]{@{}c@{}}SF\\ $i$\end{tabular}} & \textbf{\begin{tabular}[c]{@{}c@{}}ToA\\ $t_i$ (ms)\end{tabular}} & \textbf{\begin{tabular}[c]{@{}c@{}}Bitrate\\$Rb_i$ (kbps)\end{tabular}} & \textbf{\begin{tabular}[c]{@{}c@{}}Receiver Sensitivity\\ $\mathcal{S}_i$ (dBm)\end{tabular}} & \textbf{\begin{tabular}[c]{@{}c@{}}SNR threshold\\ $\psi_i$ (dB)\end{tabular}} \\ \midrule
7   & 41.22     & 5.47  & -123      & -6       \\
8   & 72.19     & 3.12  & -126      & -9       \\
9   & 144.38    & 1.76  & -129      & -12      \\
10  & 247.81    & 0.98  & -132      & -15      \\
11  & 495.62    & 0.54  & -134.5    & -17.5    \\
12  & 991.23    & 0.29  & -137      & -20      \\ \bottomrule
\end{tabular}
\end{table}

The \lora{} PHY is agnostic of higher layers.
\lorawan{} is the most widely used protocol stack for \lora{} networks.
It implements a star topology where \textit{end-devices} (nodes) connect through a single-hop to one or more \textit{gateways}, which in turn connect to a \textit{network server} via an IP network. Moreover, a \lora{} gateway can process up to nine channels in parallel, combining different sub-bands and SF~\cite{Centenaro:IEEEWC:2016}. \lora{} features the capture effect, making it possible to recover a \lora{} signal when two or more signals are received simultaneously, in the same frequency and SF, provided that the desired signal is at least $1$~dB above interference~\cite{Croce:WCL:2018}.

\subsection{\zbn{}}

\zbn{} is an amendment to the \textsc{IEEE 802.15.4} standard focusing on Smart Utility Networks (SUN) that plays an important role in the smart grid~\cite{IEEE802154g:2012}.
The standard specifies several modes operating in different bands, including the Sub-GHz ISM bands used by \lorawan{}. Multi-Rate FSK (MR-FSK), with 2-FSK or 4-FSK, is the predominant modulation version in SUN applications due to its communication range~\cite{Oh:TCE:2014}.
In this configuration, the transceiver combines FSK modulation with Frequency Hopping Spread Spectrum (FHSS) to increase robustness~\cite{Harada:ITC:2017}.
Data rate varies from $2.4$ to $200$ kbps, depending on region and frequency band.

The mandatory configuration for all regions is 2-FSK operating at $50$~kbps, which implies a channel spacing of $200$~kHz.
Transmit power depends on regional regulations, but must be at least $-3$dBm~\cite{Chang:2012}. Most configurations use $14$dBm transmit power and $1\%$ duty cycle~\cite{Munoz:Sensors:2018}.
\zbn{} also extends the MAC mechanisms defined by \textsc{IEEE 802.15.4e} amendment~\cite{IEEE802154e:2011} to make extensive use of low-energy modes.
\zbn{} networks are expected to form multi-hop, mesh networks.
Before sending data, the MAC performs either carrier sense or a simplified version of channel monitoring named Coordinated Sampled Listening (CSL)~\cite{Deshpande:ICN:2007}.

\section{System Model}
\label{sec:model}

\begin{figure}[tb]
    \centering
    \includegraphics[width=.9\columnwidth]{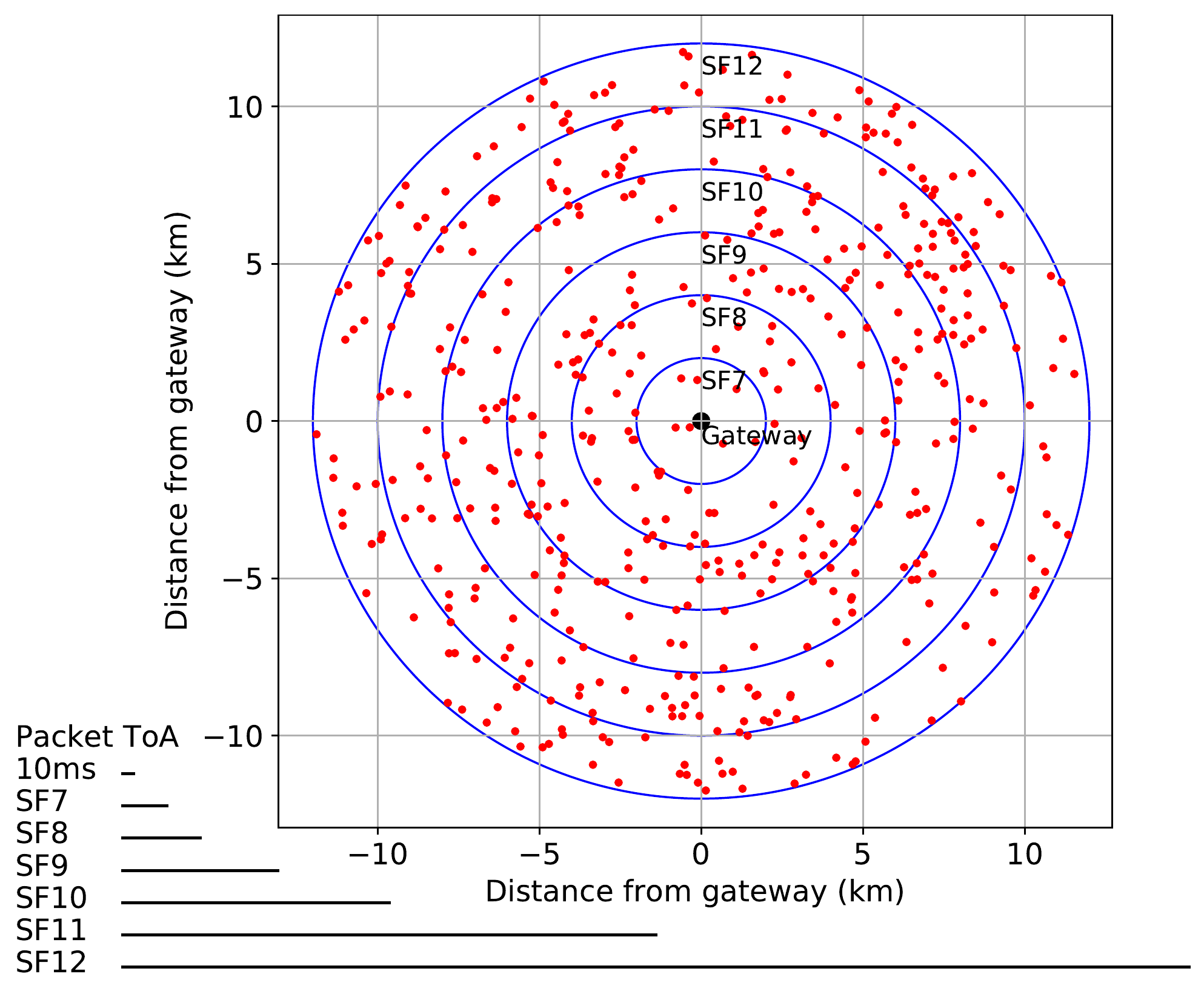}
    \caption{$\bar{N}=500$ nodes uniformly distributed in a circular area of radius $R=12000$~m around the gateway and with increasing SF every $2000$~m. The ToA, as in Table~\ref{tab:lora_specs}, is illustrated in the lower-left corner.}
    \label{fig:nodes}
\end{figure}

Following the developments in~\cite{Hoeller:Access:2018} and~\cite{Mahmood:2019}, we use a set of Poisson Point Processes (PPP)~\cite{Haenggi:Book:2012}.
Our model considers nodes deployed uniformly in a circular region around a gateway.
Figure~\ref{fig:nodes} illustrates a \emph{possible} setup where SF increases according to the distance from the gateway.
The figure, as in previous related work, uses fixed-width SF rings which are sub-optimal, an issue that is addressed by us in Section~\ref{sec:reliability}.
The vector $L=[l_0,\ldots,l_6], l_0=0,$ defines the limits of each SF ring.
Note that $R=l_6$ is the maximum network communication range, \textit{i.e.}, the coverage radius.
\lorawan{} devices transmit in the uplink at random using the ALOHA protocol and transmit once in a given period $T$.
Considering that all nodes run the same application, the network usage is different for each SF because of different data rates (see ToA in Table~\ref{tab:lora_specs}). Figure~\ref{fig:nodes} also shows the ToA difference graphically.
Hence, we model the transmission probability of \lorawan{} devices as a vector $p = [p_1, \ldots, p_6], p_i \in (0,1]~ \forall i \in \{1, \ldots, 6\}$, and $p_i = \frac{t_i}{T}$, where $t_i$ is the ToA for SF of ring $i$.
Note that, for the sake of simplicity, we define the set $S=\{1,\ldots,6\}$ to denote the SF rings and that each ring uses a respective SF in $\{7,\ldots,12\}$.

Each \lorawan{} SF ring constitutes a separated PPP, denoted $\Phi_i, i \in S$, making it possible to attribute different densities to each SF.
$\Phi_i$ has density $\alpha_i = p_i \rho_i$ in its area $V_i = \pi (l_i^2 - l_{i-1}^2)$, where $l_{i-1}$ and $l_i$ are, respectively, the inner and outer radii of SF ring $i$ (from $L$), and $\rho_i$ is the spatial density of nodes in $V_i$.
The average number of nodes in $\Phi_i$ is $\bar{N_i} = \rho_i V_i$.
The average number of nodes in the \lorawan{} network is $\bar{N} = \sum_{i \in S} \rho_i V_i$.
For instance, take the ring $i=2$ in Figure~\ref{fig:nodes}, defined by two circles of radii $l_1=2$km and $l_2=4$km. Nodes in this ring use $SF_8$.
The ring area is $V_2=\pi(l_2^2 - l_1^2)=37.7$km$^2$.
If there are, on average, $\bar{N}_2=V_2\rho_2=100$ nodes in this ring, then its spatial density is $\rho_2=\frac{\bar{N}_2}{V_2}=2.65$ nodes/km$^2$.
Finally, if nodes transmit probability is $1\%$ ($p_2=0.01$), the intensity of $\Phi_2$ is $\alpha_2=p_2\rho_2=0.0265$.

In addition to \lorawan{}, we consider an external \zbn{} network operating in the same ISM bands and geographic region. Users of that network are spread across the \lorawan{} area. \zbn{} transceivers in ISM sub-GHz bands employ bandwidth and transmit power configurations similar to \lorawan{}. So, we model the \zbn{} network as an additional PPP where nodes transmit with probability $p_z \in (0,1]$ in area $V_z = \pi R_z^2$, where $R_z \geq R$.
The PPP $\Phi_z$ has density $\alpha_z = p_z \rho_z$, where $\rho_z$ is the spatial density of nodes in $V_z$.
The average number of nodes in $\Phi_z$ is $\bar{N_z} = \rho_z V_z$.

In our analysis, $d_k$ is the Euclidean distance between the $k$-th node and the gateway, and $d_1$ denotes the distance of the node of interest.
All nodes use the same transmit power $\mathcal{P}_t$ to send signal $s_k$, while both path loss and Rayleigh fading affect the received signals of \lorawan{} and \zbn{}.
Path loss follows $g(d_k) = \left (\frac{\lambda}{4 \pi d_k}  \right )^\eta$, with wavelength $\lambda$, path loss exponent $\eta \geq 2$, while $k$ represents a device in either network. Finally, $h_k$ denotes the Rayleigh fading.
Therefore, a \lorawan{} signal $r_1$ received at the gateway is the sum of the attenuated transmitted signal $s_1$, interference, and noise, 
\begin{align}
    r_1 &= g(d_1) h_1 s_1 + \mathcal{I}_L + \mathcal{I}_Z + n,
\end{align}
where 
\begin{align}
    \mathcal{I}_L &= \sum_{i \in S} \sum_{k \in \Phi_i} {g(d_k) h_k s_k}
\end{align}
accounts for intra-network interference, considering both co-SF and inter-SF interference by summing all other received signals from all SFs, and
\begin{align}
    \mathcal{I}_Z &= \sum_{k \in \Phi_z} g(d_k) h_k s_k
\end{align}
models external interference arising, in our case, from all active nodes in the \zbn{} network.
Finally, $n$ is the additive white Gaussian noise (AWGN) with zero mean and variance $\mathcal{N} = -174 + F + 10 \textup{log}_{10}(B)$~dBm, where $F=6$dB is the receiver noise figure and $B=125$kHz is the \lora{} channel bandwidth.
The remainder of this section uses this model to derive a reliability model of \lorawan{}.

\subsection{Coverage Probability}
\label{sec:coverage}

The \textit{coverage probability} is the probability that the selected node is in coverage (not in outage), \textit{i.e.}, it can successfully communicate in the presence of noise, internal interference, and external interference.
The coverage probability of the desired node located $d_1$ meters from the gateway is thus
\begin{align}
    C_1(d_1) &= H_1(d_1) Q_1(d_1) Z_1(d_1), \label{eqn:c1} 
\end{align}
where $H_1$, $Q_1$, and $Z_1$ are described in the following sections and denote the success probability with regards to, respectively, noise, internal interference, and external interference.

\subsection{Outage Condition 1: Disconnection}
\label{sec:outage1}

Following~\cite{Hoeller:Access:2018}, we consider the disconnection probability, which depends on the communication distance.
A node is \emph{not} connected to the gateway if the SNR of the received signal is below the threshold that allows successful detection in the absence of interference.
Receiver sensitivity is different for each SF, what results in different SNR reception thresholds defined in
\vspace{.1cm}
\begin{tabular}{clllllll}
                     & $SF_7$  & $SF_8$  & $SF_9$  & $SF_{10}$  & $SF_{11}$  & $SF_{12}$ \\
    $\Psi_{[dB]} = $ & $[-6$    & $-9$    & $-12$   & $-15$      & $-17.5$    & $-20$     ],
\end{tabular}
\vspace{.1cm}\\
where $\Psi$ is the SNR threshold vector, and $\psi_i$ denotes the $i$-th element of $\Psi$, \textit{i.e.}, the SNR threshold for SF ring $i$.
Then, we model the \textit{connection probability} as
\begin{align}
    H_1(d_1) = \mathbb{P}[\textup{SNR} \geq \psi_i | d_1]. \label{eqn:h1_1}
\end{align}

Since we assume Rayleigh fading, the instantaneous SNR is exponentially distributed~\cite{Goldsmith:Book:2005}, and therefore
\begin{align}
    H_1(d_1) &= \mathbb{P} \left [ \frac{\mathcal{P}_t |h_1|^2 g(d_1)}{\mathcal{N}} \geq \psi_i \right ]
    = \textup{exp} \left ( - \frac{\mathcal{N} \psi_i}{\mathcal{P}_t g(d_1)} \right ) \label{eqn:h1}.
\end{align}

\subsection{Outage Condition 2: Intra-Network Interference}

Intra-network interference arises from the activity of other devices in the same network.
We follow~\cite{Mahmood:2019} to model both co-SF and inter-SF interference.
To recover a packet, the signal-to-interference ratio (SIR) at the gateway must be above a given threshold.
The transceiver manufacturer informs that SFs are orthogonal and that the co-SF SIR threshold is $+6$dB~\cite{Semtech:SX1272:2015}.
Goursaud \textit{et al.}~\cite{Goursaud:EAIIOT:2015} propose theoretical SIR thresholds that match Semtech co-SF value but show that different SFs are not entirely orthogonal. However, Croce \textit{et al.}~\cite{Croce:WCL:2018} showed, experimentally, that the SIR thresholds for Semtech SX1272 \lora{} transceiver are lower with regards to co-SF interference ($\approx +1$dB) but significantly higher with respect to (w.r.t.) inter-SF interference.
In this paper, we assume the experimental SIR thresholds of~\cite{Croce:WCL:2018}
\begin{equation}
\Delta_{[dB]} =
\begin{matrix}
 ~ \\
 \textup{SF}_7 \\
 \textup{SF}_8 \\
 \textup{SF}_9 \\
 \textup{SF}_{10} \\
 \textup{SF}_{11} \\
 \textup{SF}_{12} \\
\end{matrix}
\begin{matrix}
    \begin{matrix}
        \textup{SF}_7 & \textup{SF}_8 & \textup{SF}_9 & \textup{SF}_{10} & \textup{SF}_{11} & \textup{SF}_{12}
    \end{matrix} \\
    \begin{bmatrix}
       +1 &  -8 &  -9 &   -9 &   -9 &   -9 \\
     -11 &   +1 & -11 &  -12 &  -13 &  -13 \\
     -15 & -13 &   +1 &  -13 &  -14 &  -15 \\
     -19 & -18 & -17 &    +1 &  -17 &  -18 \\
     -22 & -22 & -21 &  -20 &    +1 &  -20 \\
     -25 & -25 & -25 &  -24 &  -23 &    +1
    \end{bmatrix}
\end{matrix}, \nonumber
\end{equation}
where $\Delta$ is the SIR threshold matrix, and $\delta_{i,j}$ denotes the element of $\Delta$ at the $i$-th line and $j$-th column, \textit{i.e.}, the SIR threshold for the desired signal using SF$_i$ and interference using SF$_j$.
Note that $i=j$ relates to the co-SF SIR while $i \neq j$ relates to inter-SF SIR.
If one takes the SF$_7$ column as an example, it shows how SF$_7$ interference affects the \lora{} signals.
Desired signals using higher SF are more robust to inter-SF interference, allowing for the decoding of \lora{} packets even if the interference power is much higher than the signal (\textit{e.g.}, $25$dB higher if the signal uses SF$_{12}$).

Following this rationale, we first use the formulations in~\cite{Mahmood:2019} to analyze the success probability considering the interference from  only one different SF$_j$.
Let
\begin{align}
    \textup{SIR}_j = \frac{|h_1|^2 g(d_1) \mathcal{P}_t}{\mathcal{I}_j},
\end{align}
where the interference received from nodes using SF$_j$ is
\begin{align}
    \mathcal{I}_j = \sum_{k \in \Phi_j} |h_k|^2 g(d_k) \mathcal{P}_t. \label{eqn:Ij}
\end{align}
Since the desired node at $d_1$ uses SF$_i$, the success probability is
\begin{align}
    P_{\textup{SIR}_j} (d_1, j) &= \mathbb{P}[\textup{SIR}_j \geq \delta_{i,j}] \nonumber \\
    &= \mathbb{E}_{\mathcal{I}_j} \left [ \mathbb{P} \left [ |h_1|^2 \geq \frac{\mathcal{I}_j \delta_{i,j}}{g(d_1)\mathcal{P}_t} \right ] \right]. \nonumber
\end{align}
Since $|h_1|^2 \sim \textup{exp}(1)$,
\begin{align}
    P_{\textup{SIR}_j} (d_1, j) &= \mathbb{E}_{\mathcal{I}_j} \left [ \textup{exp}\left (- \frac{\mathcal{I}_j\delta_{i,j}}{g(d_1)\mathcal{P}_t} \right) \right ]. \label{eqn:psirj1}
\end{align}

Note that~\eqref{eqn:psirj1} has the form of the Laplace Transform w.r.t. $\mathcal{I}_j$, where $\mathcal{L}_{\mathcal{I}_j} (s) = \mathbb{E}_{\mathcal{I}_j}[\textup{exp}(-s\mathcal{I}_j)]$, $s=\frac{\delta_{i,j}}{g(d_1)\mathcal{P}_t}$. Thus, using~\eqref{eqn:Ij} and applying the property of the sum of exponents,
\begin{align}
    P_{\textup{SIR}_j} (d_1, j) &= \mathbb{E}_{\Phi_j,|h_k|^2} \left[ \prod_{k \in \Phi_j} \textup{exp}(-s |h_k|^2 g(d_k) \mathcal{P}_t) \right]. \nonumber
\end{align}
Solving the expectation over $|h_k|^2 \sim \textup{exp}(1)$ yields
\begin{align}
    P_{\textup{SIR}_j} (d_1, j) &= \mathbb{E}_{\Phi_j} \left[ \prod_{k \in \Phi_j} \frac{1}{1 + s g(d_k) \mathcal{P}_t} \right]. \nonumber
\end{align}
We solve the expectation over the PPP $\Phi_j$ using the probability generating functional of the product over PPPs where $\mathbb{E}[\prod_{x\in\Phi_j}f(x)]=\textup{exp}[-\alpha_j\int_{\mathbb{R}^2} (1-f(x))\textup{d}x]$, with $\alpha_j$ as the density of $\Phi_j$, converting $d_k$ to polar coordinates, and replacing $s$, obtaining
\begin{align}
    P_{\textup{SIR}_j}(d_1,j) &= \textup{exp}\left( -2\pi\alpha_j \int_{l_{j-1}}^{l_j} \frac{\delta_{i,j}d_1^\eta}{x^\eta + \delta_{i,j}d_1^\eta} x ~\textup{d}x \right). \label{eqn:psirjopen} 
\end{align}

As a contribution over~\cite{Mahmood:2019}, we provide in Appendix~\ref{apx:ffunc} a closed-form solution for the integral in~\eqref{eqn:psirjopen}, which we isolate in function $f(\cdot)$.
The solution is
\begin{align}
    f(d_1, \delta_{i,j}, l_{j-1}, l_j) &= 
      \frac{l_j^2}{2} \, {}_2F_1\left(1, \frac{2}{\eta}; 1+\frac{2}{\eta}; -\frac{l_j^\eta}{d_1^\eta \delta_{i,j}}   \right) \nonumber \\
    &- \frac{l_{j-1}^2}{2} \, {}_2F_1\left(1, \frac{2}{\eta}; 1+\frac{2}{\eta}; -\frac{l_{j-1}^\eta}{d_1^\eta \delta_{i,j}}   \right), \label{eqn:fhyp}
\end{align}
and therefore,
\begin{align}
    P_{\textup{SIR}_j}(d_1,j) = \textup{exp} \left[ -2\pi\alpha_j f(d_1, \delta_{i,j}, l_{j-1}, l_j) \right].\label{eqn:psirj}
\end{align}

Now we consider interference from all SFs when
\begin{align}
    \textup{SIR} = \frac{|h_1|^2 g(d_1) \mathcal{P}_t}{\sum_{j \in S} \mathcal{I}_j}.
\end{align}
Following~\cite{Mahmood:2019}, we consider that an outage takes place if the SIR for at least one interfering SF exceeds the threshold in~$\Delta$.
Conversely, the probability that a collision does not occur is
\begin{align}
    Q_1(d_1) &= \prod_{j\in S} P_{\textup{SIR}_j} (d_1,j). \nonumber
\end{align}
Since $P_{\textup{SIR}_j}$, shown in~\eqref{eqn:psirj}, is an exponential function, we compute the above product by summing the exponents and reorganizing, obtaining
\begin{align}
    Q_1(d_1) &= \textup{exp} \left ( -2\pi \sum_{j \in S} \alpha_j f(d_1, \delta_{i,j}, l_{j-1}, l_j) \right ). \label{eqn:q1}
\end{align}

\subsection{Outage Condition 3: External Interference}

Orfanidis \textit{et al.}~\cite{Orfanidis:WiMob:2017} report the selectivity of \lora{} receivers in the presence of \zbn{} signals. We use Orfanidis \textit{et al.} experimentally obtained isolation thresholds to analyze the SIR in the presence of external interference generated by an \zbn{} network.
Here, we model the \zbn{} network as PPP $\Phi_z$ and consider~\cite{Orfanidis:WiMob:2017}
\vspace{.1cm}
\begin{tabular}{clllllll}
                        & $SF_7$  & $SF_8$  & $SF_9$  & $SF_{10}$  & $SF_{11}$  & $SF_{12}$  \\
    $\Theta_{[dB]} = $ & $[-6$    & $-9$    & $-12.5$ & $-16$      & $-16$      & $-16$      ],
\end{tabular}
\vspace{.1cm}\\
where $\Theta$ denotes the \lora{} vs. \zbn{} SIR threshold vector, and $\theta_i$ denotes the $i$-th element of $\Theta$, \textit{i.e.}, the SIR threshold for the desired signal in SF ring $i$ and interference from the \zbn{} network.

The analysis of the LoRa capture probability in the presence of \zbn{} interference is similar to the case for one SF ($P_{\textup{SIR}_j}$), but taking the $\Theta$ vector and the \zbn{} network parameters into account.
For
\begin{align}
    \textup{SIR}_z = \frac{|h_1|^2 g(d_1) \mathcal{P}_t}{\sum_{k \in \Phi_z} |h_k|^2 g(d_k) \mathcal{P}_t},
\end{align}
the \emph{capture probability} w.r.t. external interference is
\begin{align}
    Z_1(d_1) &= P_{\textup{SIR}_z} (d_1) = \textup{exp}(-2\pi \alpha_z  f(d_1, \theta_i, 0, R_z) ). \label{eqn:z1}
\end{align}
Note that the model for external interference supports any other communication technology given that adequate SIR thresholds of $\theta$ are provided.

\section{Optimum \lorawan{} Configuration}
\label{sec:reliability}

The expressions in Section~\ref{sec:model} determine the expected reliability of a single node located at a given distance from the gateway. \textit{However, what if one wants to plan the network deployment?}
In this section, we consider the use of the previous model to this end.
We first consider the inversion of the expressions in the model to obtain network configurations for a targeted minimum average reliability.
Afterward, we propose two algorithms that derive optimum network configurations supporting the desired minimum reliability requirement.

\subsection{Guaranteeing the Reliability Target}

To start our search for optimal \lorawan{} configurations we invert the previously described outage expressions defined in~\eqref{eqn:h1} and~\eqref{eqn:q1}, so the network parameters can be extracted from them to achieve a minimum desired reliability level.
Note that~\eqref{eqn:z1} does not depend on the \lorawan{} configuration.
It is, however, taken into account in the optimization algorithm proposed in Sections~\ref{sec:maxR} and~\ref{sec:maxN} to consider external interference.
One can assume that, in our network model, the nodes presenting the worst average reliability in each SF ring are those on the outer ring limit.
It happens because the signals emitted by those nodes suffer greater path-loss and are, therefore, more susceptible to interference.

\subsubsection{SF Ring Limits}

As a first step, we find the maximum distance that ensures the required minimum average reliability level w.r.t. the connection probability $H_1$. We denote this threshold by $\mathcal{T}_{H_1}$.
We rewrite~\eqref{eqn:h1} to perform operations over the SNR threshold $\psi_i$ in $\Psi$ and the outer SF ring limit $l_i$,
\begin{align}
    \mathcal{T}_{H_1} &= \textup{exp}\left ( -\frac{\mathcal{N}\psi_i}{\mathcal{P}_t g(l_i)} \right),
\end{align}
and then it is straightforward to obtain
\begin{align}
    l_i &= \frac{\lambda}{4\pi} \left[ -\frac{\mathcal{P}_t \textup{ln}(\mathcal{T}_{H_1})}{\mathcal{N}\psi_i}\right]^{\frac{1}{\eta}}. \label{eqn:h1_inv}
\end{align}
Note that the radius of the overall coverage area is $R = l_6$.

\subsubsection{Ring Densities}

Since~\eqref{eqn:h1_inv} defines the network geometry, it is now possible to obtain the outage due to external interference observed by the nodes at each ring edge from $Z_1(l_i)$.
After that, we compute the maximum densities of the PPPs in $Q_1$ that satisfy the given final reliability target $\mathcal{T}$, the previously assumed connection reliability target $\mathcal{T}_{H_1}$, and the external interference of each SF $i$.
Thus, following~\eqref{eqn:c1}, making $C_1(l_i)=\mathcal{T}$ and $H_1(l_i)=\mathcal{T}_{H_1}$, we have for each SF ring $i$ that $\mathcal{T} = \mathcal{T}_{H_1} Q_1(l_i) Z_1(l_i),$ and thus
\begin{align}
 \frac{\mathcal{T}}{\mathcal{T}_{H_1}  Z_1(l_i)} &= \textup{exp} \left( -2\pi \sum_{j \in S} \alpha_j f(l_i, \delta_{i,j}, l_{j-1},l_j) \right). \label{eqn:reliability}
\end{align}
In~\eqref{eqn:reliability}, the function $f(\cdot)$ is independent of $\alpha_j$ if the SF limits are pre-defined.
Then, let $y_{i,j}=f(l_i, \delta_{i,j}, l_{j-1}, l_j)$ and $b_i=-\frac{1}{2\pi} \textup{ln}\left(\frac{\mathcal{T}}{\mathcal{T}_{H_1} Z_1(l_i)}\right)$, yielding, for each SF ring $i$,
\begin{align}
    y_{i,1}\alpha_1 + y_{i,2}\alpha_2 + y_{i,3}\alpha_3 + y_{i,4}\alpha_4 + y_{i,5}\alpha_5 + y_{i,6}\alpha_6 = b_i. \label{eqn:eqn_syst}
\end{align}
If we name the vectors $A = [ \alpha_1, \ldots, \alpha_6 ], B = [ b_1, \ldots, b_6 ],$ and matrix $Y = [ y_{i,j} ], \forall i,j \in S$, from~\eqref{eqn:eqn_syst}, we derive a system of linear equations $Y \times A = B$ and solve it for the PPPs densities $A$ by making
\begin{align}
    A = Y^{-1} \times B \label{eqn:q1_inv}.
\end{align}
Note that $Y$ is a $|S| \times |S|$ square matrix, both $A$ and $B$ are row vectors of length $|S|$, and all values $y_{i,j}$ in $Y$ are positive real numbers.
To be invertible ($Y^{-1}$), $Y$ must have a non-zero determinant.
Considering the diagonal method to compute the determinant of $Y$, we observe that, due to $\Delta$, the values at $i=j$ are significantly higher than when $i\neq j$, thus making the positive diagonal greater than the negative diagonal, yielding a determinant that is virtually never zero.

\subsection{Maximization of Communication Range}
\label{sec:maxR}

The expressions presented above allow us to obtain twelve network parameters: six communication range limits $L=[l_1,\ldots,l_6]$ from~\eqref{eqn:h1_inv}, and six PPP densities $A=[\alpha_1,\ldots,\alpha_6]$ from~\eqref{eqn:q1_inv}.
Note that~\eqref{eqn:q1_inv} depends on~\eqref{eqn:h1_inv} because of $L$.
Combining both equations generates an incomplete linear system of six equations and twelve variables.
In order to search for optimized feasible network configurations, we propose an algorithm that uses~\eqref{eqn:h1_inv} and~\eqref{eqn:q1_inv} in an iterative method, trying to extend the outer SF ring limits as much as possible, while preserving the targeted final reliability level $\mathcal{T}$ and ensuring service to a minimum quantity of nodes ($N_{min}$).
The algorithm extends the outer SF ring limits by reducing $\mathcal{T}_{H_1}$.
Similarly, increasing $\mathcal{T}_{H_1}$ shortens these limits.
The algorithm iteratively guesses values for $\mathcal{T}_{H_1}$ and then, after obtaining $L$ through~\eqref{eqn:h1_inv}, analyzes the maximum possible densities $A$.
As $\mathcal{T}_{H_1}$ gets closer to $\mathcal{T}$, the capture probability $Q_1$ increases and, with fixed $L$ and $Z_1$, higher $Q_1$ is possible only with lower densities.
It may lead to configurations breaking the $N_{min}$ restriction.
Conversely, if $\mathcal{T}_{H_1}$ is too close to $1$, the outer limits of the SF rings will be shorter, leading to small coverage areas that are  useless in practice.
However, the proposed algorithm identifies feasible ranges for the network parameters, thus dealing with this parameter trade-off.

Algorithm~\ref{alg:max_range} employs a bisection technique~\cite{Burden:Book:2016} to explore the network design space (\textit{i.e.}, possible values of $\mathcal{T}_{H_1}$), seeking to maximize the width of each SF ring and, as a consequence, the network communication range (disk radius), while preserving the targeted minimum reliability $\mathcal{T}$ and ensuring service to, at least, a given number of nodes ($N_{min}$).
The bisection technique fits well to our problem because it accelerates convergence by reducing the design space in half in each iteration, and it is guaranteed to converge if the problem is feasible.
The inputs of the algorithm are the targeted reliability $\mathcal{T}$, the duty-cycle vector $p$, $N_{min}$, and the density of \zbn{} interfering nodes $\alpha_z$.
The algorithm outputs a $result$ variable stating if the algorithm converged ($1$) or not ($-1$), the achieved number of nodes $N$, and the vectors $L$ and $A$ containing, respectively, the rings limits and densities ensuring the target reliability.

\begin{algorithm}
    \caption{Maximization of SF rings widths given the target reliability ($\mathcal{T}$) and the minimum number of nodes ($N_{min}$).\label{alg:max_range}}
    \begin{algorithmic}[1]
        \renewcommand{\algorithmicrequire}{\textbf{Input:}}
        \renewcommand{\algorithmicensure}{\textbf{Output:}}
        \REQUIRE $\mathcal{T}, p, N_{min}, \alpha_z$
        \ENSURE $\textup{result}, A, L, N$

        \STATE $\mathcal{T}_{H_1,max} \leftarrow 1$ \\ 
        \STATE $\mathcal{T}_{H_1,min} \leftarrow \mathcal{T}$
        \STATE $R \leftarrow 0, N \leftarrow 0$
        \STATE $\textup{result} \leftarrow 0$

        \WHILE {$\textup{result} = 0$}
            \STATE $\mathcal{T}_{H_1} \leftarrow (\mathcal{T}_{H_1,max} + \mathcal{T}_{H_1,min})/2$
            \STATE $L \leftarrow \frac{\lambda}{4\pi} \left[ -\frac{\mathcal{P}_t \textup{ln}(\mathcal{T}_{H_1})}{\mathcal{N}\Psi}\right]^{\frac{1}{\eta}}$ \COMMENT{Equation \eqref{eqn:h1_inv}}
            \STATE $R_{last} \leftarrow R$
            \STATE $R \leftarrow L[end]$
            \STATE $R_z \leftarrow R$
            \FOR{$i=[1, \ldots, 6]$}
                \FOR{$j=[1, \ldots, 6$]}
                    \STATE $Y[i,j] \leftarrow f(l_i, \delta_{i,j}, l_j, l_{j+1})$
                \ENDFOR
                \STATE $B[i] \leftarrow -\frac{1}{2\pi}\ln{\frac{\mathcal{T}}{Z_1(l_i) \mathcal{T}_{H_1}}}$
            \ENDFOR
            \STATE $A \leftarrow Y^{-1} \times B$  \COMMENT{Equation \eqref{eqn:q1_inv}}
            \STATE $V \leftarrow \textup{ComputeRingAreas(L)}$
            \STATE $N \leftarrow (A  p) \times V'$

            \IF{abs($R$ - $R_{last}$) $< \chi$ ~\AND~ $A_i \geq 0, \forall A_i$}
                \IF{$N < N_{min}$}
                    \IF{$\mathcal{T}_{H_1,max}-\mathcal{T}_{H_1,min} < \epsilon$}
                        \STATE $\textup{result} \leftarrow -1$
                    \ENDIF
                \ELSE
                    \STATE $\textup{result} \leftarrow 1$
                \ENDIF
            \ENDIF
            \IF{result = 0}
                \IF{$A_i \geq 0, \forall A_i$ \AND $N \geq N_{min}$}
                    \STATE $\mathcal{T}_{H_1,max} = \mathcal{T}_{H_1}$
                \ELSE
                    \STATE $\mathcal{T}_{H_1,min} = \mathcal{T}_{H_1}$
                \ENDIF
            \ENDIF
        \ENDWHILE
        \RETURN $\textup{result}, A, L, N$
    \end{algorithmic}
\end{algorithm}

After initializing the variables (lines 1-4), the optimization loop starts and runs until the algorithm converges (line 26) or diverges (line 23).
The optimization procedure  ``guesses'' values for $\mathcal{T}_{H_1}$, trying to reduce it to enlarge the width of each SF ring, thus increasing the coverage area.
Note that since $C_1$ depends on $H_1$ from~\eqref{eqn:c1}, $\mathcal{T}_{H_1}$ must be greater than $\mathcal{T}$; otherwise, both $Q_1$ and $Z_1$ would have to be $1$, which is impossible in practice.
Thus, Algorithm~\ref{alg:max_range} sets the initial search region for $\mathcal{T}_{H_1}$ to $(\mathcal{T},1)$.
The guessed value for $\mathcal{T}_{H_1}$ in each iteration is at the center of this region, as expressed in line 6.
At each iteration, if the selected $\mathcal{T}_{H_1}$ generates a configuration where the number of nodes is above $N_{min}$, it is assumed that $Q_1$ can be enhanced by decreasing $N$, which allows for further decreasing $\mathcal{T}_{H_1}$ in the next iteration.
Conversely, if $N<N_{min}$, $\mathcal{T}_{H_1}$ is increased so that $Q_1$ can be lower, allowing for more nodes in the network.
This ``binding'' part of the algorithm is in lines 30-34.

Provided the branch-and-bound technique guesses $\mathcal{T}_{H_1}$ in line 6, the range limits for all SF are computed using~\eqref{eqn:h1_inv} in line 7.
In the following, the algorithm uses the newly computed vector $L$ to obtain vector $B$ and matrix $Y$ (lines 11-16), allowing for the computation of the PPPs densities in $A$ (line 17), using~\eqref{eqn:q1_inv}.
Following that, the number of nodes fitting the generated configuration is computed by first obtaining the area of each SF ring in lines 18-19 as $V_i = \pi (l_i^2 - l_{i-1}^2)$.
The algorithm converges and stops when the difference in the radius $R$ of the overall coverage area between two consecutive iterations is less than $\chi$ (line 20) and $N > N_{min}$ (line 21), where $\chi$ defines the precision of $R$.
If the variation of $R$, \textit{i.e.}, $\textup{abs}(R-R_{last})$, is too small and the algorithm did not achieve $N_{min}$ yet, the algorithm keeps trying to converge until the variation in the guessed $\mathcal{T}_{H_1}$ is below a threshold $\epsilon$ (line 22), in which case the algorithm stops and announces a divergence.
After evaluating the proposed algorithm for a set of test scenarios, we concluded that good values for the stopping thresholds are $\chi=1m$ and $\epsilon=10^{-9}$.

Algorithm~\ref{alg:max_range} always converges if there are feasible solutions to the problem.
If the requirements of minimum network density ($N_{min}$ and $p$), targeted reliability ($\mathcal{T}$), or both, are too high, however, the algorithm may take too long to converge. 
Hence, we stop the algorithm when the changes in $\mathcal{T}_{H_1}$ get too small (line 22), thus guaranteeing that the algorithm will not run forever since $\mathcal{T}_{H_1,max} - \mathcal{T}_{H_1,min}$ decreases every iteration.
It is important to note, however, that achieving higher network density is always possible by reducing the reliability requirement $\mathcal{T}$.
Moreover, highly demanding scenarios without feasible solutions or with lengthy convergence are not typical in \lorawan{}, since the technology has been conceived to support massive rather than critical IoT applications.

The algorithm has linear complexity, \textit{i.e.}, $\mathcal{O}(n)$.
Analysis of convergence time of this method depends on the precision, which we define in Algorithm 1 as $\chi=1$ meter.
Therefore, the literature defines the maximum number of iterations as $n=\textup{log}_2\left(\frac{\chi_0}{\chi}\right)$, where $\chi_0$ is the initial error~\cite{Burden:Book:2016}.
In Algorithm 1, $\chi_0 = |R_1 - R_0|$, with $R_0=0$ (Algorithm 1, line 3) and $R_1$ computed in the first iteration (line 9) using (18) with $\mathcal{T}_{H_1} = \frac{1+\mathcal{T}}{2}$ (line 6).
Note that although Algorithm 1 involves the solution of a system of linear equations with matrix inversion (line 17) and matrix multiplications (line 19), these are computed quite efficiently since it handles low-dimension matrices -- the largest matrix is \textbf{Y}, which is 6-by-6.

\subsection{Maximization of Number of Nodes}
\label{sec:maxN}

Now we consider the case of maximizing the total number of nodes given restrictions of minimum coverage radius ($R_{min}$) and average reliability ($\mathcal{T}$).
We use a more straightforward approach than in Algorithm~\ref{alg:max_range}.
The problem of maximizing the number of nodes is equivalent to the problem of minimizing $Q_1$.
Thus it is straightforward to conclude, from~\eqref{eqn:c1}, that we should maximize $H_1$ because higher $H_1$ allows for lower $Q_1$.
Since we assume that the worst cases are at the edge of the SFs and we have a restriction on the coverage radius, the maximum possible $H_1$ is that yielding $l_6 = R_{min}$.
Thus, from~\eqref{eqn:h1}, we conclude that $\mathcal{T}_{H_1} = H_1(R_{min})$.
Assuming the same $\mathcal{T}_{H_1}$ for all SFs, we use~\eqref{eqn:h1_inv} to compute $L$ and obtain the geometry of the network.

Once we obtain $L$, we get the maximum allowable densities ensuring $\mathcal{T}$ through~\eqref{eqn:q1_inv} as shown in Algorithm~\ref{alg:max_nodes}.
Line 1 uses~\eqref{eqn:h1} to compute the maximum $\mathcal{T}_{H_1}$ satisfying $R_{min}$.
Line 2 uses the computed $\mathcal{T}_{H_1}$ to obtain the geometry of the network $L$.
The loop in lines 5-10 computes matrix $Y$ and vector $B$, so the maximum device density vector $A$ can be computed in line 11.
Finally, after computing the areas of the rings and storing them in vector $V$ (line 12), we obtain the maximum number of nodes in line 13.

\begin{algorithm}
    \caption{Maximization of the number of nodes given the target reliability ($\mathcal{T}$) and the minimum coverage radius ($R_{min}$).\label{alg:max_nodes}}
    \begin{algorithmic}[1]
        \renewcommand{\algorithmicrequire}{\textbf{Input:}}
        \renewcommand{\algorithmicensure}{\textbf{Output:}}
        \REQUIRE $\mathcal{T}, p, R_{min}, \alpha_z$
        \ENSURE $\textup{result}, A, L, N_{max}$

        \STATE $\mathcal{T}_{H_1} \leftarrow \textup{exp}\left( -\frac{\mathcal{N}\psi_6}{\mathcal{P}_t g(R_{min})}\right)$ \\
        \STATE $L \leftarrow \frac{\lambda}{4\pi} \left[ -\frac{\mathcal{P}_t \textup{ln}(\mathcal{T}_{H_1})}{\mathcal{N}\Psi}\right]^{\frac{1}{\eta}}$ \COMMENT{Equation (1)}

        \STATE $R \leftarrow L[end]$
        \STATE $R_z \leftarrow R$
        \FOR{$i=[1, \ldots, 6]$}
            \FOR{$j=[1, \ldots, 6$]}
                \STATE $Y[i,j] \leftarrow f(l_i, \delta_{i,j}, l_j, l_{j+1})$
            \ENDFOR
            \STATE $B[i] \leftarrow -\frac{1}{2\pi}\ln{\frac{\mathcal{T}}{Z_1(l_i) \mathcal{T}_{H_1}}}$
        \ENDFOR

        \STATE $A \leftarrow Y^{-1} \times B$  \COMMENT{Equation (2)}
        \STATE $V \leftarrow \textup{ComputeRingAreas(L)}$
        \STATE $N_{max} \leftarrow (A  p) \times V'$

        \IF {$A_i \geq 0, \forall A_i$}
            \STATE $\textup{result} \leftarrow 1$
        \ELSE
            \STATE $\textup{result} \leftarrow -1$
        \ENDIF

        \RETURN $\textup{result}, A, L, N_{max}$
    \end{algorithmic}
\end{algorithm}

Note that Algorithm~\ref{alg:max_nodes} is not iterative since there is no loop searching for the optimum solution and, therefore, it has complexity $\mathcal{O}(1)$.
This algorithm merely describes how to use the proposed models to determine the optimum \lorawan{} configuration considering the restrictions.
The approach produces unfeasible configurations if the restrictions are too strict.
Thus, lines 14-18 check whether the method generated non-negative densities for all SFs to assess whether the results are feasible or not.

\section{Numerical Results}
\label{sec:results}

This section evaluates the proposed model and algorithms.
In all figures, lines represent theoretical probabilities (\textit{i.e.}, $H_1, Q_1, Z_1, Q_1$), while marks along the lines show the results of Monte Carlo simulations.
Each mark in a figure is the average of $10^5$ simulations considering random deployments.
Moreover, we assume $F=6$~dB, $\eta=2.75$, $\lambda=c/f$~m, $c=3\times 10^8$~m/s (speed of light), $f=868$~MHz for both \lorawan{} and \zbn{}.
\lorawan{} channel bandwidth is $B_l=125$~kHz, and \zbn{} channel bandwidth is $B_z=200$~kHz.
We also assume that nodes in \lorawan{} and \zbn{} transmit with $\mathcal{P}_t=14$dBm.
These parameters configure typical European sub-urban scenarios.

Concerning \zbn{} interference, we evaluate the algorithms considering three scenarios.
In real deployments, the designer of a \lorawan{} network may not know the operational parameters of the interfering \zbn{} network. Thus, in a practical situation, the designer should assume worst-case configurations for the external network.

\subsection{Model Validation}

\begin{figure}[tb]
    \centering
    \includegraphics[width=0.9\columnwidth]{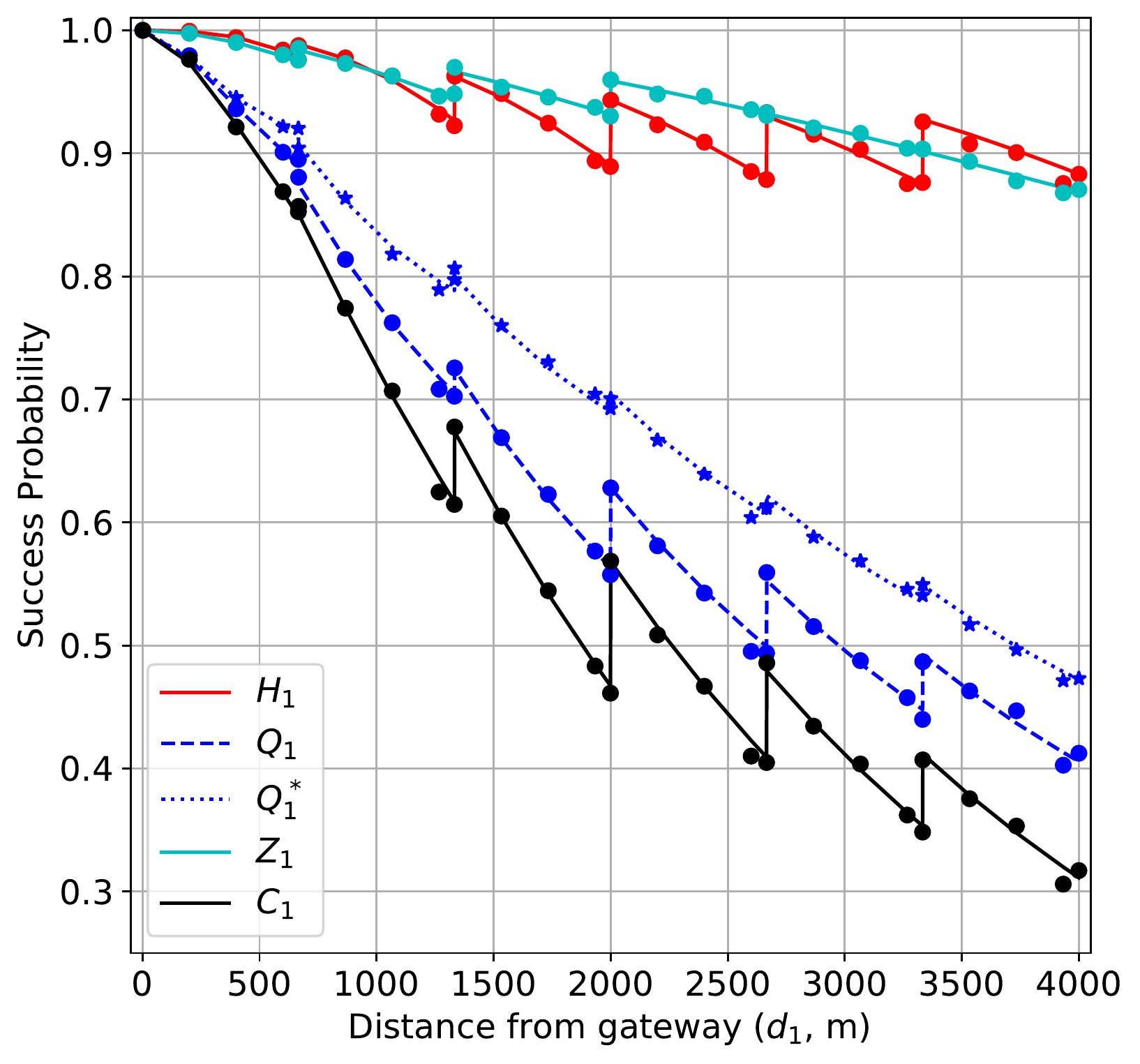}
    \caption{Success probabilities of all outage sources. \lora{}: $\bar{N}=4000$, $p=0.1\%$, $\eta=2.75$, $\mathcal{P}_t=14$dBm, $R=4000m$. \zbn{}: $\bar{N}_z=1000$, $p_z=0.1\%$, $\eta=2.75$, $\mathcal{P}_t=14$dBm, $R_z=4000m$.}
    \label{fig:model_vs_sim}
\end{figure}

Figure~\ref{fig:model_vs_sim} aims to validate the presented models by showing the success rates $H_1$, $Q_1$, $Z_1$, and $C_1$ as a function of the distance from the gateway.
The scenario considers an average number of nodes $\bar{N}=4000$, transmitting with duty cycle $p=0.1$\% in a circular area around the gateway with radius $R=4000$m.
The \zbn{} network generating external interference has $\bar{N}_z=1000$ nodes transmitting with duty cycle $p_z=0.1$\%, also in a circular area with radius $R_z=4000$m.
As can be seen, all theoretical expressions (lines) match the simulation results (marks).
One can observe in $Z_1$ that a relatively light interference from \zbn{} ($\bar{N_z}=1000, p_z=0.1\%$) has little impact in lower SFs due to the smaller ToA and reduced probability of concurrent transmissions.
Higher SFs, on the other hand, have higher ToA and thus suffer more from this external interference.

Also, in Figure~\ref{fig:model_vs_sim}, $Q_1^*$ shows what the capture probability would be if we consider that \lora{} signals are perfectly orthogonal.
We obtain $Q_1^*$ from~\eqref{eqn:q1} by considering only $j=i$.
As can be seen, the gap between $Q_1$ and $Q_1^*$ shows that inter-SF interference plays a vital role in link quality.

\subsection{Algorithm 1: Maximization of Range}

\begin{figure}[tb]
    \centering
    \includegraphics[width=.95\columnwidth]{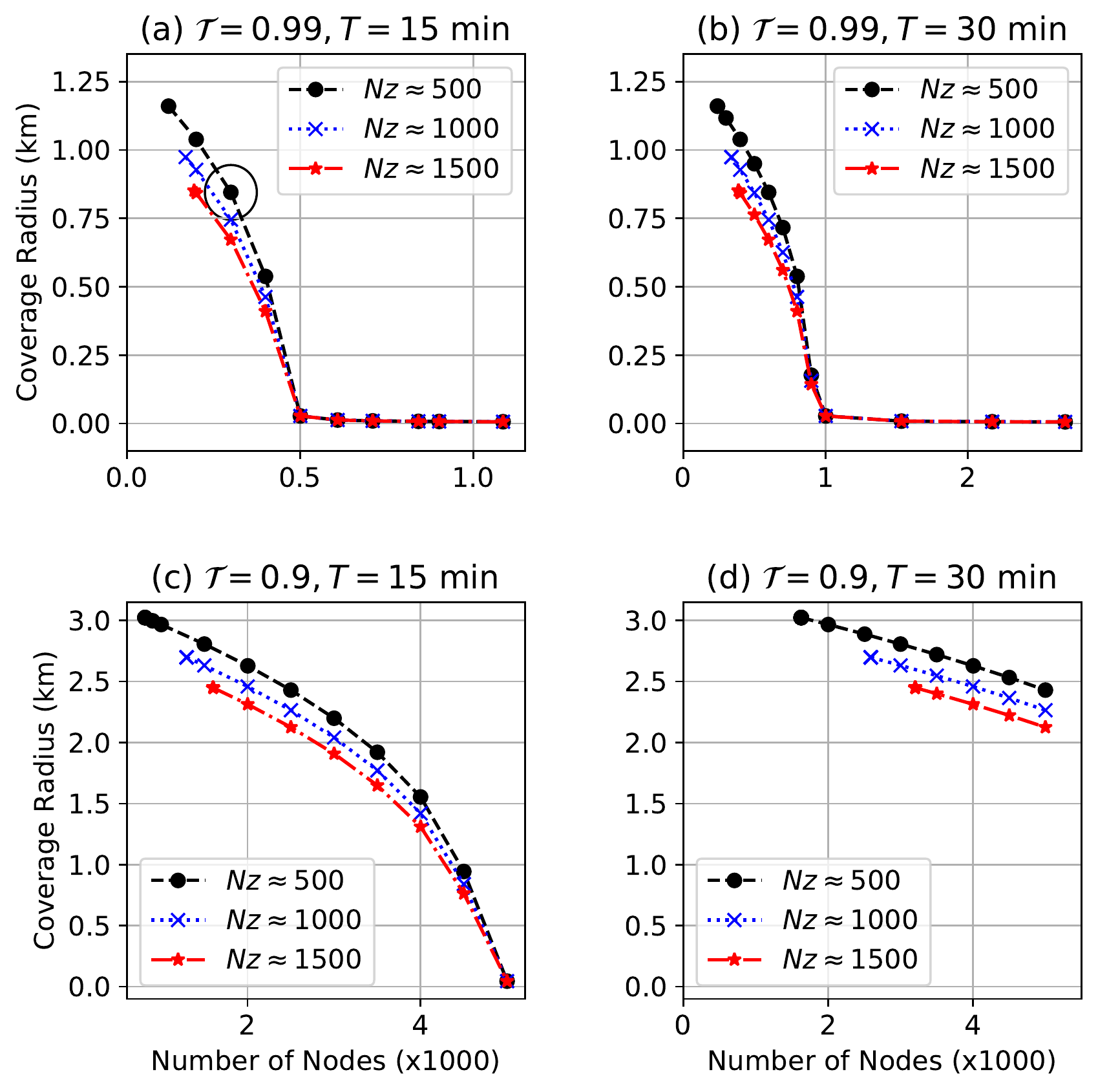}
    \caption{Optimization between coverage and number of nodes given a minimum reliability constraint when maximizing $R$ with Algorithm~\ref{alg:max_range}.}
    \label{fig:maxR}
\end{figure}

Now we evaluate Algorithm~\ref{alg:max_range} of Section~\ref{sec:maxR}.
These results use the same network parameters employed to validate the network model.
Figure~\ref{fig:maxR} presents a series of graphs for varying optimization objectives.
Plots in the same row consider the same reliability target $\mathcal{T}$, while plots in the same column use the same packet generation interval $T$, expressed in minutes.
Each graph shows three curves, each one considering a different amount of \zbn{} interference, which varies by changing the number of \zbn{} nodes ($\bar{N}_z$), always with duty cycle $p_z=0.1\%$.
Each optimization point considers different $N_{min}$ values, evaluated every $100$m.

The first conclusion, when comparing the curves in all plots, is that different \zbn{} interference leads to shorter communication ranges when following our proposed optimization procedure.
That makes sense since shorter distances feature smaller path loss, making signals less susceptible to external interference.
It is also possible to observe that less stringent reliability targets lead to larger coverage areas.
Again, that makes sense since smaller $\mathcal{T}$ yields smaller $\mathcal{T}_{H_1}$, which in turn enables longer communication range.

Also, in Figure~\ref{fig:maxR}, plot~(a) shows the more rigorous scenario;
the configuration allowing the required reliability is only practical for $N_{min} \leq 400$ nodes, with a radius varying from $410$ to $1160$ meters, depending on $N_{min}$ and the external interference.
The coverage radius with $N_{min} \geq 500$ either converged to unpractical distances of less than $10$ meters or diverged, meaning that we could not place this many \lorawan{} nodes  with packet generation interval of $15$ minutes while ensuring minimum reliability of $\mathcal{T}=0.99$.

For $\mathcal{T}=0.99$, there are more feasible scenarios if network usage decreases.
Plot~\ref{fig:maxR}b shows that configurations with up to $900$ nodes are possible if the packet transmission interval is $T=30$ minutes.
For $\mathcal{T}=0.9$ with $T=15$ minutes, it is possible to find reasonably good network configurations up to $N_{min}=4500$.
However, $N_{min}=5000$ shrinks the communication range to unpractical distances.

\begin{figure}
    \centering
    \includegraphics[width=.95\columnwidth]{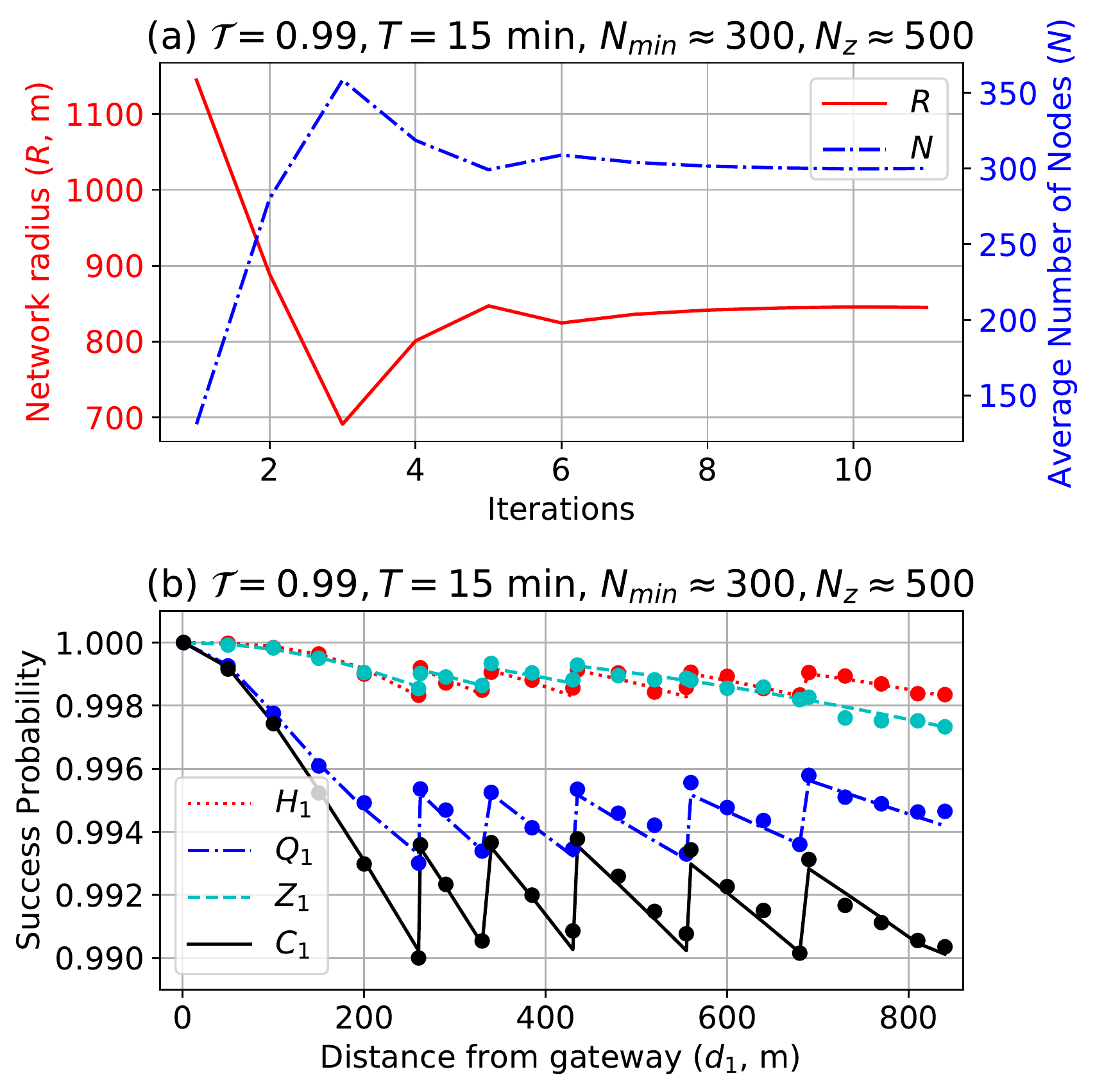}
    \caption{Convergence of Algorithm~\ref{alg:max_range} and success probability for the scenario marked in Figure~\ref{fig:maxR}a.}
    \label{fig:conv_model}
\end{figure}

Figure~\ref{fig:conv_model} illustrates the behavior of Algorithm~\ref{alg:max_range} and network performance when taking the circled case in Figure~\ref{fig:maxR}a as an example.
Figure~\ref{fig:conv_model}a shows the convergence of $R$ and $\bar{N}$ for this scenario.
Applying the estimate of the number of iterations presented in Section~\ref{sec:maxR} to this example makes $R_1=1244.7$m because of $\mathcal{T}=0.99$.
Therefore, the maximum number of iterations to reach $|R-R_{last}|<\chi$ (line 20, Algorithm~\ref{alg:max_range}) is $n=\textup{log}_2\left(\frac{\chi_0}{\chi}\right)=\textup{log}_2\left(\frac{1244.7}{1}\right) = 10.28$. We see in Figure~\ref{fig:conv_model}a that the algorithm converges after the 11th iteration with $\bar{N}=300.1$ , thus ``stretching'' the network range as much as possible.
Also, note that convergence time depends on $\chi_0$, which in turn depends on $\mathcal{T}$.
If we consider the same case above with $\mathcal{T}=0.9$ or $\mathcal{T}=0.8$, we would have, respectively, $\chi_0=2899.7$ or $\chi_0=3767.3$, what would make, respectively, $n=11.50$ or $n=11.88$, showing that the impact of $\mathcal{T}$ in convergence time is small.

Table~\ref{tab:alg_conv} shows numerical results of the same scenario in two columns: ``All sources'' with the results for our complete model; and ``Intra-SF only'' disregarding both inter-SF and external interference sources.
We get the results in the ``Intra-SF only'' column using the same models, but setting $\theta_i = -\infty, \forall i \in S$ in $\Theta$, and $\delta_{i,j} = 1$ for $i=j$ and $\delta_{i,j}=-\infty$ otherwise.
When considering all sources of interference, as expected, the fact that ToA impacts the duty cycle induces the optimization procedure to allocate most of the nodes with lower SF.
That happens because signal attenuation increases with distance, making more distant nodes more vulnerable to both internal and external interference.
Recall that a shorter ToA reduces the collision probability.
Moreover, longer ToA generates more internal interference to other SFs.

In some cases, higher SFs may not be used to ensure minimum reliability.
However, note that \eqref{eqn:h1_inv}, \eqref{eqn:q1_inv}, and Algorithm~\ref{alg:max_range} can be extended to change the restriction $N_{min}$ to represent a vector with the minimum number of nodes using each SF.
One can achieve that by revisiting the computation of the densities in~\eqref{eqn:q1_inv} to consider such a minimum number of nodes when computing the spatial density.
Since doing that will possibly result in more nodes using higher SFs, it is expected that fewer nodes use lower SF, resulting in a smaller total number of nodes, as well as a shorter network radius, since the algorithm will converge to a higher $H_1$ to compensate the increased $Q_1$.
When disregarding inter-SF and external interference, we observe that higher SFs are profoundly affected by inter-SF and external interference, mainly due to their extended ToA.
In particular, we observe that interference, rather than path loss, is the main factor for which our method disfavors the use of higher SF.
Moreover, it is clear that interference considerably affects coverage.

\begin{table}[tb]
\caption{Detailed optimization results for the marked scenario in Figure~\ref{fig:maxR}a.}
\label{tab:alg_conv}
\centering
\begin{tabular}{@{}cc|ccc|ccc@{}}
\toprule
\multicolumn{2}{c|}{Interference:} & \multicolumn{3}{c|}{All sources}    & \multicolumn{3}{c}{Intra-SF only}   \\ \midrule
\multirow{2}{*}{Scenario} & \multirow{2}{*}{SF} & Range & \multirow{2}{*}{$\bar{N}_i$} & \multirow{2}{*}{$\bar{N}$} & Range & \multirow{2}{*}{$\bar{N}_i$} & \multirow{2}{*}{$\bar{N}$} \\
                          &                     &  (m)  &             &           &  (m)  &             &           \\ \midrule
\multirow{6}{*}{\ref{fig:maxR}a} 
                     & 7  & 261.6   & 162.8   & \multirow{6}{*}{300.1}  &  370.0  & 124.6   & \multirow{6}{*}{300.0} \\
                     & 8  & 336.4   &  67.5   &                         &  475.7  &  87.1   &                         \\
                     & 9  & 432.4   &  32.5   &                         &  611.6  &  43.5   &                         \\
                     & 10 & 555.9   &  21.8   &                         &  786.2  &  25.3   &                         \\
                     & 11 & 685.4   &  10.6   &                         &  969.3  &  12.9   &                         \\
                     & 12 & 845.0   &   4.7   &                         & 1195.1  &   6.4   &                \\ \bottomrule
\end{tabular}
\end{table}

Finally, Figure~\ref{fig:conv_model}b shows the success probabilities of the example scenario, where the optimized configurations consider the minimum average reliability target $\mathcal{T}$ for all distances from the gateway.
As expected, the success probability approaches the desired minimum $\mathcal{T}=0.99$ at the edge of each SF.
We can see that collisions ($Q_1$) are kept almost constant or increase slightly with SF. That happens because the algorithm reduces the number of nodes using each SF to keep $Q_1$ in pace with $H_1$ and $Z_1$, to ensure the minimum  $\mathcal{T}$.

\subsection{Algorithm~2: Maximization of Nodes}

\begin{figure}[tb]
    \centering
    \includegraphics[width=.95\columnwidth]{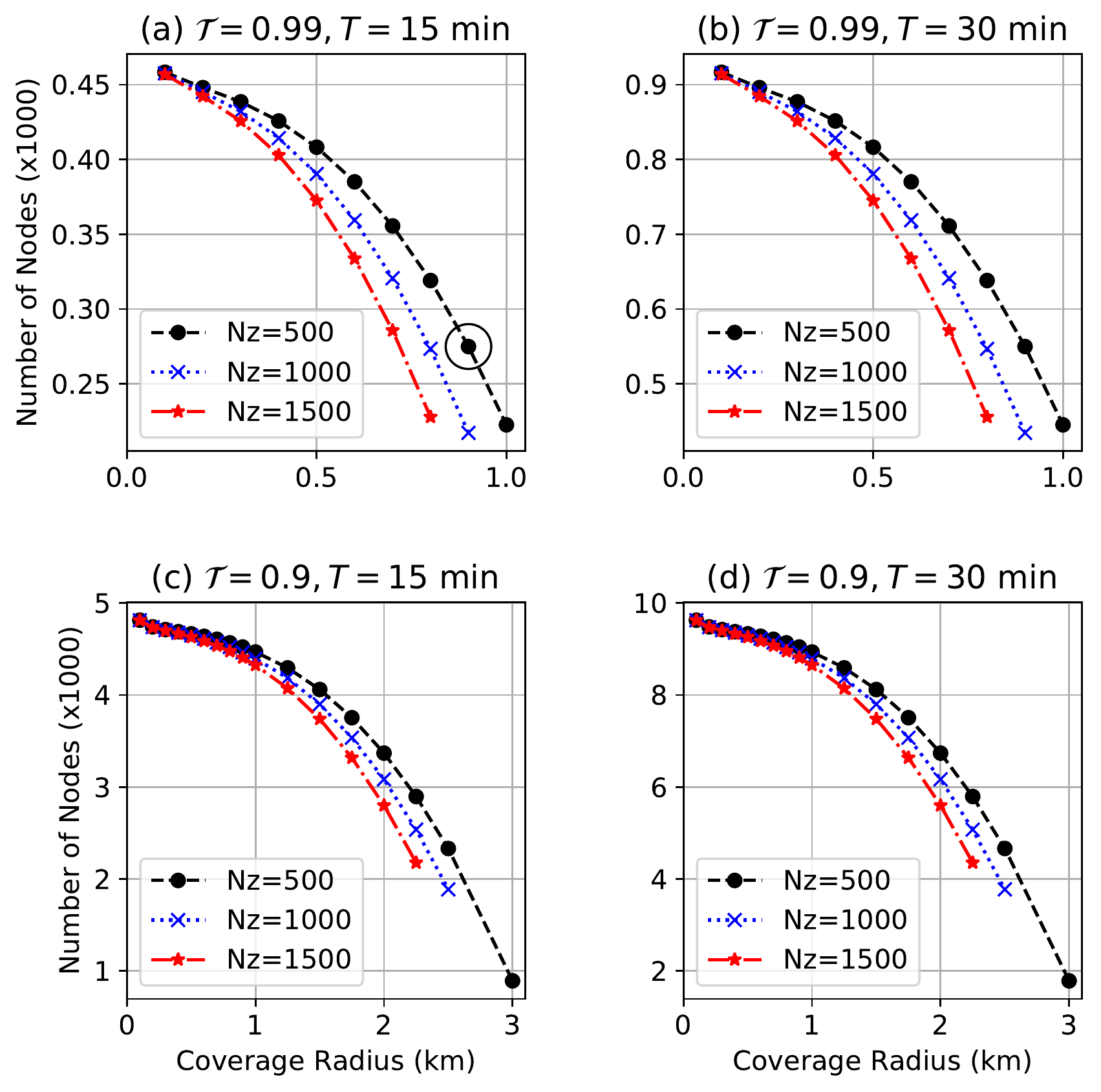}
    \caption{Optimization between the number of nodes and coverage given a minimum reliability constraint when maximizing $\bar{N}$ with Algorithm~\ref{alg:max_nodes}.}
    \label{fig:maxN}
\end{figure}

Now we evaluate Algorithm~\ref{alg:max_nodes} of Section~\ref{sec:maxN}.
The plots in Figure~\ref{fig:maxN} show the results for different scenarios of required minimum reliability ($\mathcal{T}$) and message generation period ($T$).
For all plots, the x-axis represents the $R_{min}$ input to the algorithm, while the y-axis shows the achieved maximized number of nodes.
In each plot, the x-axis grows up to the value for which the requirements yield practical results.

In Figure~\ref{fig:maxN}, if we analyze each row of plots independently, we see that the maximum number of nodes is a linear function of the transmission period $T$.
For instance, considering $R_{min}=500$m and $N_z=500$, $N_{max}$ in plots~\ref{fig:maxN}a and~\ref{fig:maxN}b are, respectively, $408.18$ and $816.36$, \textit{i.e.}, $N_{max}$ doubles when $T$ doubles.
That is expected since these variations ensure the same network load in all scenarios.
We also observe, in all plots, that increased external interference reduces both the number of nodes and the achievable coverage radius.

\begin{table}[tb]
\caption{Detailed optimization results for the marked scenario in Figure~\ref{fig:maxN}a.}
\label{tab:algN_conv}
\centering
\begin{tabular}{@{}cc|ccc|ccc@{}}
\toprule
\multicolumn{2}{c|}{Interference:} & \multicolumn{3}{c|}{All sources}    & \multicolumn{3}{c}{Intra-SF only}   \\ \midrule
\multirow{2}{*}{Scenario} & \multirow{2}{*}{SF} & Range & \multirow{2}{*}{$\bar{N}_i$} & \multirow{2}{*}{$\bar{N}$} & Range & \multirow{2}{*}{$\bar{N}_i$} & \multirow{2}{*}{$\bar{N}$} \\
                          &                     &  (m)  &             &           &  (m)  &             &           \\ \midrule
\multirow{6}{*}{\ref{fig:maxN}a}
                     & 7  & 278.7   & 149.2   & \multirow{6}{*}{274.9}  &  278.7  & 211.1   & \multirow{6}{*}{508.2} \\
                     & 8  & 358.3   &  61.8   &                         &  358.3  & 147.6   &                         \\
                     & 9  & 460.6   &  29.9   &                         &  460.6  &  73.7   &                         \\
                     & 10 & 592.1   &  20.2   &                         &  592.1  &  42.9   &                         \\
                     & 11 & 730.0   &   9.5   &                         &  730.0  &  21.8   &                         \\
                     & 12 & 900.0   &   4.0   &                         &  900.0  &  10.9   &                \\ \bottomrule
\end{tabular}
\end{table}

\begin{figure}[tb]
    \centering
    \includegraphics[width=.95\columnwidth]{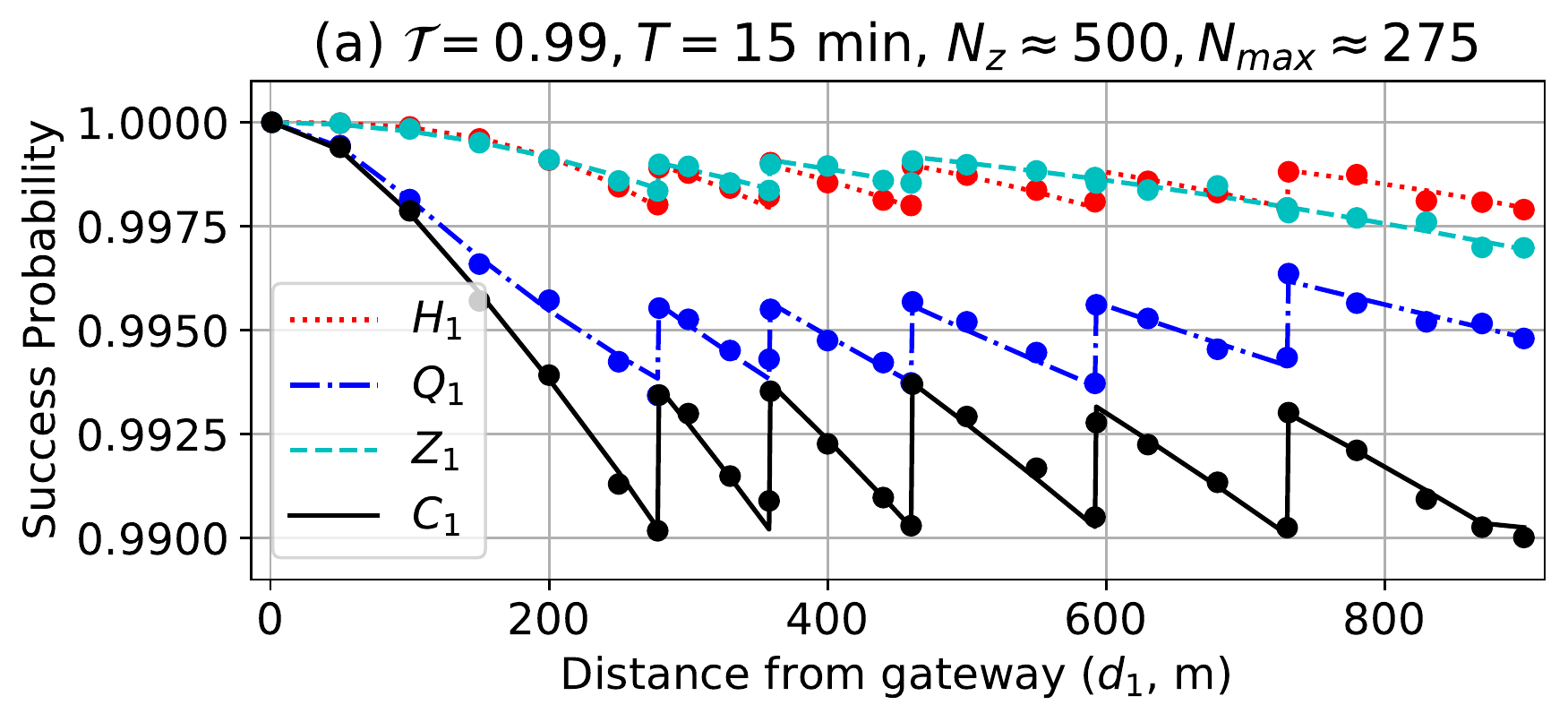}
    \caption{Average outage expectation for the marked scenario in Figure~\ref{fig:maxN}a.}
    \label{fig:maxN-perf}
\end{figure}

Table~\ref{tab:algN_conv} shows the achieved geometry and number of nodes of the marked scenario of Figure~\ref{fig:maxN}a.
Again, the ``All sources'' column presents the results of our complete model, while the ``Intra-SF only'' column disregards inter-SF and external interference.
Since the method assumes that the maximum number of nodes is achieved with the shortest possible distances, the maximum range of a node using SF$_{12}$ has to be $R_{min}$ ($900$m for this case).
As for Algorithm~\ref{alg:max_range}, Algorithm~\ref{alg:max_nodes} also favors lower SFs.
Moreover, Table~\ref{tab:algN_conv} shows that the maximum number of nodes almost doubles when disregarding inter-SF and external interference, emphasizing the importance of taking such impairments into account to avoid overestimating the network performance. Finally, Figure~\ref{fig:maxN-perf} shows success probabilities for the example scenario, which approach $\mathcal{T}=0.99$ at SF edges but stay above the required minimum $\mathcal{T}$ for all distances shorter than $R_{min}$.

\section{Conclusion}
\label{sec:conclusion}

This paper presents two algorithms to optimize the configuration of \lorawan{} under imperfect SF orthogonality and \zbn{} interference.
We use models of \lorawan{} networks to derive success probabilities of packet delivery under internal and external (\zbn{}) interference.
The presented algorithms search for optimum \lorawan{} configurations given restrictions of minimum network density or coverage radius, meeting a target minimum reliability level.
The analytic results are validated using simulations.

Regarding \zbn{} interference over \lorawan{}, although higher SF should be more robust to this type of interference, they suffer more from that impairment because their increased ToA makes it more likely that transmissions overlap with \zbn{} activity. Finally, regarding the proposed algorithms, they provide a tool for exploring trade-offs between network load and coverage range by showing the feasible region of \lorawan{} network configurations.

Possible future extensions of this work can include other important LPWAN features, such as power allocation and multiple base stations. Moreover, one may consider adapting the proposed algorithms to maximize reliability given fixed network geometry and number of users, in a way similar to what~\cite{Jia:TVT:2017} does, or to analyze latency and energy efficiency of the models, similarly to~\cite{Mukherjee:Netw:2018}, noting that neither \cite{Jia:TVT:2017} nor \cite{Mukherjee:Netw:2018} consider \lorawan{} networks.
In the future, we also plan to validate our models using data from a multi-technology large scale IoT network at the University of Oulu.

\section*{Appendix A: Solution of $f(d_1,\gamma,l_a,l_b)$}\label{apx:ffunc}

Here we solve the integral in~\eqref{eqn:psirjopen}.
Let
\begin{align}
    f(d_1, \gamma, l_a, l_b) = \int_{l_a}^{l_b} \frac{\gamma d_1^\eta}{x^\eta + \gamma d_1^\eta} x ~\textup{d}x. \nonumber
\end{align}
We rearrange $f(\cdot)$ as
\begin{align}
    f(d_1, \gamma, l_a, l_b) &= \int_{l_a}^{l_b} x \left(\frac{x^\eta}{d_1^\eta \gamma} + 1\right)^{-1} \textup{d}x  \nonumber
\end{align}
and use the binomial theorem $(x+1)^{-1}=\sum_{k=0}^{\infty}(-1)^k x^k$ to obtain
\begin{align}
    f(d_1, \gamma, l_a, l_b) &= \int_{l_a}^{l_b} \sum_{k=0}^{\infty} \left(\frac{-1}{d_1^\eta \gamma}\right)^{k} x^{\eta k +1}  \textup{d}x. \nonumber
\end{align}
Since $f(\cdot)$ is continuous in $\mathbb{R} \, \forall x>0$, we interchange the sum and the integration and solve the integral, yielding
\begin{align}
    f(d_1, \gamma, l_a, l_b) &= \sum_{k=0}^{\infty}  \left(\frac{-1}{d_1^\eta \gamma}\right)^{k} \frac{x^{2+\eta k}}{\eta k +2} \Biggr|_{l_a}^{l_b}. \nonumber
\end{align}
We resort to the Pochhammer function $(a)_k = a(a+1)\cdots(a+k-1) = \frac{\Gamma(a+k)}{\Gamma(a)}$ and to $\frac{(b)_k}{(b+1)_k} = \frac{b}{b+k}$, and reorganize $f(\cdot)$ as 
\begin{align}
    f(d_1, \gamma, l_a, l_b) &= \frac{x^2}{2} \sum_{k=0}^{\infty} \frac{(1)_k}{k!} \frac{\left(\frac{2}{\eta}\right)_k}{\left(1+\frac{2}{\eta}\right)_k} \left(-\frac{x^\eta}{d_1^\eta \gamma} \right)^k  \Biggr|_{l_a}^{l_b}, \nonumber
\end{align}
which is in the form of the Gaussian Hypergeometric function $_2F_1(a,b;c;z) = \sum_{k=0}^{\infty} \frac{(a)_k (b)_k}{(c)_k}\frac{z^k}{k!}$~\cite{Daalhuis:Chapter:2010}, what yields
\begin{align}
    f(d_1, \gamma, l_a, l_b) &= \frac{x^2}{2} \, {}_2F_1\left(1, \frac{2}{\eta}; 1+\frac{2}{\eta}; -\frac{x^\eta}{d_1^\eta \gamma}   \right) \Biggr|_{l_a}^{l_b} . \label{eqn:fhyp} \nonumber
\end{align}

\balance

\bibliographystyle{IEEEtran}
\bibliography{main}

\end{document}